# Extended Kalman Filter State Estimation for Autonomous Competition Robots

Ethan Kou[1] and Acshi Haggenmiller[2#]
[1] Henry M Gunn High School, Palo Alto, CA, USA
[2] University of Michigan    [#] Advisor
https://github.com/BubblyBingBong/EKF

Autonomous mobile robot competitions judge based on a robot's ability to quickly and accurately navigate the game field. This means accurate localization is crucial for creating an autonomous competition robot. Two common localization methods are odometry and computer vision landmark detection. Odometry provides frequent velocity measurements, while landmark detection provides infrequent position measurements. The state can also be predicted with a physics model. These three types of localization can be "fused" to create a more accurate state estimate using an Extended Kalman Filter (EKF). The EKF is a nonlinear full-state estimator that approximates the state estimate with the lowest covariance error when given the sensor measurements, the model prediction, and their variances. In this paper, we demonstrate the effectiveness of the EKF by implementing it on a 4-wheel mecanum-drive robot simulation. The position and velocity accuracy of fusing together various combinations of these three data sources are compared. We also discuss the assumptions and limitations of an EKF.

## 1. INTRODUCTION

The First Tech Challenge (FTC) is a robotics competition hosted by the non-profit organization called For Inspiration and Recognition of Science and Technology (FIRST). The competition field (Figure 1) is a flat area measuring 12 ft x 12. Obstacles may be present to increase difficulty. The field is enclosed by a wall, which has multiple navigation images taped to it. Images can be detected by the robot's camera.

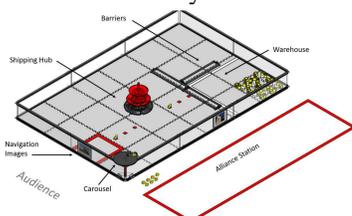

Figure 1:  2021-2022 FTC Season Competition Field [1]

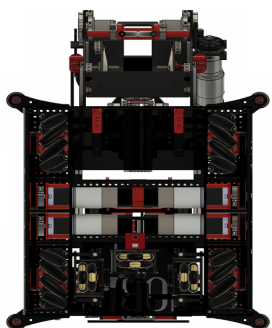

Figure 2: Robot bottom view: The robot being described in this paper drives with four mecanum wheels (wheels with small rollers attached at a 45-degree angle). The top of the figure is the front of the robot. There are three dead-wheel encoders (wheels that track how far they spun) on the drivetrain of the robot, two vertical and one horizontal. There is also a camera attached to the top of the robot.

A competition match in FTC consists of two parts: a two-minute driver-controlled period and a thirty-second autonomous period. Points are scored by transporting objects from one location to another. In the 2021-2022 FTC challenge, points are scored by transporting objects from the warehouse to the shipping hub (Figure 1). This makes a robot's ability to accurately traverse the field crucial to scoring points. Odometry and computer vision landmark detection are localization methods that each have benefits and drawbacks. Odometry provides new data hundreds of times per second and measures velocity, but error in velocity will accumulate in position. Computer vision provides position data only when a wall image is in the robot's camera view.

An Extended Kalman Filter can be used to combine odometry's frequent updates with landmark detection's position accuracy to improve the state estimate. The Extended Kalman Filter is a full-state estimator, meaning that it can estimate state values even if they aren't being directly measured. This can be used to give an accurate position estimate, even when a wall image is not in the camera's view.

In this paper, we demonstrate the effectiveness of the Extended Kalman Filter for the mecanum drive robot shown in Figure 2, using a Python simulation.

## 2. STATISTICS PRIMER

### 2.1. Gaussian Distributions

Let X be a continuous random variable. A Gaussian distribution is the same as a normal distribution, which is a probability density function defined by the variable's mean μ and standard deviation σ. In a Gaussian distribution, E(X) = μ. The Gaussian distribution density function has equation (1) and its graph is Figure 3.

$$f(x) = \frac{1}{\sigma\sqrt{2\pi}} e^{-\frac{1}{2}\left(\frac{x-\mu}{\sigma}\right)^2} \qquad (1)$$

For all Gaussian Distributions, the probability that a randomly generated number is in the interval [x1, x2]

is $\int_{x1}^{x2} f(x)\,dx$. The total area under a Gaussian distribution is always 1.

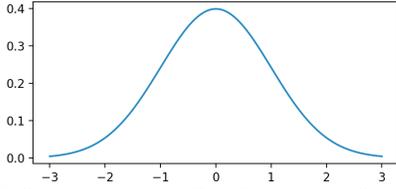

Figure 3: Standard Normal Distribution: μ = 0 and σ = 1

## 2.2. Variance and Covariance

The variance of a Gaussian random variable X is:

$$Var(X) = E[(X - E[X])^2] = \sigma^2 \qquad (2)$$

The covariance matrix of an n x 1 multivariate Gaussian is an n x n diagonal matrix such that the element in column and row i equals the variance of the Gaussian in index i of the multivariate Gaussian. For 2 Gaussians, X and Y, the following equations hold true:

$$E[X + Y] = E[X] + E[Y] \qquad (3)$$
$$Var(X \pm Y) = Var(X) + Var(Y) \qquad (4)$$

The variance formula only holds if X and Y are independent.

## 3. LOCALIZATION

### 3.1. The Kalman Filter

*1) System Model*

Kalman Filters (KF) are widely used in the robotics industry, especially for localization. This is because a KF can fuse multiple types of localization data to create a more accurate estimate. The KF operates in three steps. First, the KF predicts the state using a mathematical model. Then, it measures the state with sensors. Finally, it uses the measurement to correct the predicted state to produce the estimated state. It is assumed that the prediction, measurement, and estimation are all multivariate Gaussians which are characterized by a mean vector and a covariance matrix. Noise is a multivariate Gaussian distribution with μ = 0. The KF calculates the estimate by performing a weighted average between the prediction and measurement, where lower variances are given higher weight. The KF is a perfect linear state estimator because the estimation has the minimum possible error covariance. Figure 4 illustrates the state estimation process.

It is also assumed that the system can be modelled linearly. In a linear model, the transformation from the state to the prediction and measurement must be linear combinations of the state values. A linear model can't multiply state values or take the sines/cosines of state values.

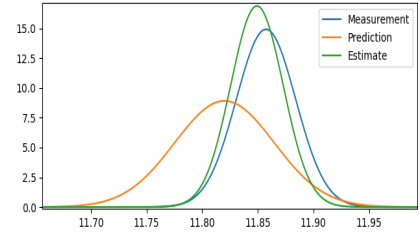

Figure 4: Visualization of state estimation: The distributions of prediction, measurement, and estimation are shown. The estimation is weighted towards the measurement because it has a lower variance.

The prediction and measurement are:

$$x_{a,t} = Ax_{a,t-1} + Bu_t + q_t \qquad (5)$$

(5) is the prediction/dynamics model. $x_{a,t}$ (n x 1) represents the true state at time step t. $u_t$ (k x 1) is the control vector at time step t, which is used to move the robot. A (n x n) is the state transition matrix, and B (n x k) is the control matrix. $q_t$ (n x 1) is prediction Gaussian noise with covariance matrix Q. In the linear Kalman Filter, A and B are constant. In our simulation, we add artificial noise $q_t$ when simulating the robot's true state since on a physical robot, there will always be some randomness.

$$z_t = Hx_{a,t} + r_t \qquad (6)$$

(6) is the measurement model. $z_t$ (m x 1) represents the measured state at time t. H (m x n) is the state-to-measurement matrix, which picks out the values in the state that are being measured because H is a matrix of 0s and 1s. Gaussian noise $r_t$ (m x 1) with covariance matrix R is measurement noise.

*2) State Estimation*

$x_{a,t}$ and $z_t$ are predicted based on the most recent belief/estimation. The estimated state at time step t is denoted as $x_t$. Subscript p denotes prediction.

$$x_p = Ax_{t-1} + Bu_t \qquad (7)$$
$$z_p = Hx_p \qquad (8)$$

The mean value is predicted as it has the highest probability of being the actual value. The covariance matrix of the state x is P, which is predicted with the following equation:

$$P_p = AP_{t-1}A^T + Q \qquad (9)$$

The covariance matrices are used to compute the Kalman Gain, which is denoted as K (n x m):

$$K = P_p H^T \left(H P_p H^T + R\right)^{-1} \qquad (10)$$

The Kalman Gain determines how much the measurement should be trusted by utilizing the variance of the measurement, the prediction, and the latest estimation. If the Kalman Gain is higher, the estimate will be weighted more towards the measurement. Finally, The Kalman Gain is used to compute the final estimation.

$$x_t = x_p + K(z_t - z_p) \qquad (11)$$

$$P_t = P_p - KHP_p \qquad (12)$$

*3) Limitations of Linearity*

The KF requires a completely linear model, which has many limitations.

The robot state contains the robot's heading, and since the ut vector represents movement in body coordinates and the position state is in global coordinates, a rotation must be present in the prediction model. A rotation requires taking the sines/cosines of the robot's angle state, making the system nonlinear. As a result, we must use the Extended Kalman Filter to handle the nonlinearity of the system.

### 3.2. The Extended Kalman Filter

The Extended Kalman Filter (EKF) performs state estimation on a nonlinear system, but it isn't optimal. The reason for its suboptimal estimation will be explained shortly. The EKF equations are very similar to the KF equations.

$$x_{a,t} = f(x_{a,t-1}, u_t) + q_t \qquad (13)$$

$$z_t = h(x_{a,t}) + r_t \qquad (14)$$

$$x_p = f(x_{t-1}, u_t) \qquad (15)$$

$$z_p = h(x_p) \qquad (16)$$

where f is the state transition function and h is the state-to-measurement function. Both can be nonlinear. To calculate the Kalman gain, which uses the same equation as the KF, we obtain linear matrix approximations of functions f and h with Jacobians A and H respectively.

$$A = \frac{\partial f}{\partial x_{t-1}} \qquad (17)$$

$$H = \frac{\partial h}{\partial x_p} \qquad (18)$$

A Jacobian is a matrix of partial derivatives. If we let $x_i$ be the $i^{th}$ element of the function's input and $y_i$ be the $i^{th}$ element of the function's output, the element in the Jacobian at row r and column c is $\frac{\partial y_r}{\partial x_c}$. This linear approximation is what makes an EKF non-optimal. Jacobians A and H also need to be recalculated at each time step. For more KF and EKF resources, visit [2], [3], and [4].

## 4. SIMULATION SYSTEM MODEL

### 4.1. System Overview

We will be using the EKF because the robot's system is nonlinear. The Δt of the true state is 0.001s and the Δt of the EKF is 0.01s. All variables used in the mathematics for the prediction model and measurement model are labelled in Figure 5, which also illustrates the coordinate system. Instead of denoting the robot's state as xt, we denote it as st. This is done to avoid confusion between the state and the x position. The robot's state st consists of its x, y, and θ positions as well as its velocity in body coordinates. Body velocity is used instead of global velocity because body velocity tends to be less nonlinear, resulting in improved localization accuracy.

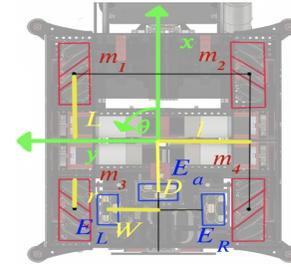

Figure 5: The diagram is a top view of the robot's drivetrain. The robot is x-forwards and y-left. Angle (radians) increases when rotating counter clockwise. Since the FTC field is based off inches, the position state is in inches and the velocity state is in inches/second. The red outlines are mecanum wheels, and the direction of the rollers are shown. The blue outlines are encoders for odometry. EL and ER are vertical encoders (turn when moving forward/backward), and Ea is a horizontal encoder (turn when moving left/right). Yellow lines represent constant lengths.

Including a subscript for a matrix or vector means that the subscript applies to all elements in the matrix/vector if applicable.

$$s_t = [x \quad y \quad \theta \quad \dot{x} \quad \dot{y} \quad \dot{\theta}]_t^T \qquad (19)$$

The control vector contains the velocities of the four mecanum wheels in radians per second:

$$u_t = [m_1 \quad m_2 \quad m_3 \quad m_4]^T \qquad (20)$$

A positive mecanum wheel velocity means that the wheel will spin in a way to move the robot forwards. Since $u_t$ contains velocity, the robot's acceleration can be unrealistically large. This problem can be solved by adding acceleration to the state $s_t$, which won't be done in this paper.

The robot receives six measurements in total: three from odometry and three from landmark detection. The odometry measurements are the dead-wheel encoder velocities (inches per second) and are denoted as EL, ER, and Ea respectively. Landmark detection provides three position measurements, xm, ym, and $\theta_m$. Measurement zt is written as:

$$z_t = [x_m \quad y_m \quad \theta_m \quad E_L \quad E_R \quad E_a]^T \quad (21)$$

Odometry measurements are provided at every time step, while landmark detection measurements are only provided when an image landmark is completely in the robot's view. The camera field of view (FOV) is 70 degrees. When an image is not in sight, the position measurement variance is set to a very high value to make the EKF put negligible weight towards the camera measurement.

### 4.2. Prediction Model

*1) State Transition Function*

The state transition function f propagates the current state to the next state. To find f, we need to express the next state in terms of the current state. The dynamics model used to calculate the true state for the simulation is simply the prediction model with added noise qt at each time step.

The next position state is predicted by turning body velocity into global velocity:

$$\begin{bmatrix} x \\ y \\ \theta \end{bmatrix}_p = \begin{bmatrix} x + \Delta t(\dot{x}\cos\theta - \dot{y}\sin\theta) \\ y + \Delta t(\dot{x}\sin\theta + \dot{y}\cos\theta) \\ \theta + \Delta t \cdot \dot{\theta} \end{bmatrix}_{t-1} \quad (22)$$

The next velocity state depends solely on the $u_t$ vector. Using mecanum wheel mechanics [5], we can calculate the robot's body velocity based on the velocity of each of the four mecanum wheels. L is the distance from the center horizontal axis to the center of a mecanum wheel and l is the distance from the center vertical axis to the center of a mecanum wheel. r is the radius of a mecanum wheel. Refer to Figure 5.

$$\begin{bmatrix} \dot{x} \\ \dot{y} \\ \dot{\theta} \end{bmatrix}_p = \left( \frac{r}{4} \begin{bmatrix} 1 & 1 & 1 & 1 \\ 1 & -1 & -1 & 1 \\ -\frac{1}{L+l} & \frac{1}{L+l} & -\frac{1}{L+l} & \frac{1}{L+l} \end{bmatrix} \right) \begin{bmatrix} m_1 \\ m_2 \\ m_3 \\ m_4 \end{bmatrix}_t \quad (23)$$

Multiplying the right side, we get:

$$\begin{bmatrix} \dot{x} \\ \dot{y} \\ \dot{\theta} \end{bmatrix}_p = \frac{r}{4} \begin{bmatrix} m_1 + m_2 + m_3 + m_4 \\ m_1 - m_2 - m_3 + m_4 \\ \frac{1}{L+l}(-m_1 + m_2 - m_3 + m_4) \end{bmatrix}_t \quad (24)$$

Combining (22) and (24), we have:

$$f\left(\begin{bmatrix} x \\ y \\ \theta \\ \dot{x} \\ \dot{y} \\ \dot{\theta} \end{bmatrix}_{t-1}, \begin{bmatrix} m_1 \\ m_2 \\ m_3 \\ m_4 \end{bmatrix}_t\right) = \begin{bmatrix} x_{t-1} + \Delta t(\dot{x}\cos\theta - \dot{y}\sin\theta)_{t-1} \\ y_{t-1} + \Delta t(\dot{x}\sin\theta + \dot{y}\cos\theta)_{t-1} \\ \theta_{t-1} + \Delta t \cdot \dot{\theta}_{t-1} \\ \frac{r}{4}(m_1 + m_2 + m_3 + m_4)_t \\ \frac{r}{4}(m_1 - m_2 - m_3 + m_4)_t \\ \frac{r}{4(L+l)}(-m_1 + m_2 - m_3 + m_4)_t \end{bmatrix} \quad (25)$$

Noise qt has covariance matrix Q:

$$Q = diag$$
$$([0.002\Delta t \quad 0.002\Delta t \quad 0.002\Delta t \quad 0.45 \quad 0.45 \quad 0.45]) \quad (26)$$

diag() transforms a 1 x n vector into an n x n diagonal matrix. The position variance is multiplied by $\Delta t$ because error increases with more delay between updates. The velocity variance, however, is independent of the value of $\Delta t$ These variances are not obtained from measurements on an actual robot. They are just arbitrary values that can be tweaked to adjust the simulation behaviour. For example, when experimenting solely with the accuracy of the prediction model, the values in Q can be set to near zero, which will cause the KF to "ignore" the measurement.

*2) State Transition Jacobian*

Based on (25), we can compute Jacobian matrix A.

$$A = \begin{bmatrix} 1 & 0 & \Delta t(-\dot{x}\sin\theta - \dot{y}\cos\theta) & \Delta t \cdot \cos\theta & -\Delta t \cdot \sin\theta & 0 \\ 0 & 1 & \Delta t(\dot{x}\cos\theta - \dot{y}\sin\theta) & \Delta t \cdot \sin\theta & \Delta t \cos\theta & 0 \\ 0 & 0 & 1 & 0 & 0 & \Delta t \\ 0 & 0 & 0 & 0 & 0 & 0 \\ 0 & 0 & 0 & 0 & 0 & 0 \\ 0 & 0 & 0 & 0 & 0 & 0 \end{bmatrix}_{t-1} \quad (27)$$

### 4.3. Measurement Model

*1) State-to-measurement Function*

The state-to-measurement function h converts the current state into the current predicted measurement. The model for the true measurement includes added noise rt. In order to obtain the h function, we need to express the predicted measurement in terms of the predicted state.

This is simple for the position components as computer vision directly measures position. Measurement is denoted by the subscript m.

$$\begin{bmatrix} x \\ y \\ \theta \end{bmatrix}_m = \begin{bmatrix} x \\ y \\ \theta \end{bmatrix}_p \quad (28)$$

To obtain encoder readings in terms of predicted velocity, we use the linear odometry equations (29) [6].

$$\begin{bmatrix} \dot{\theta} \\ \dot{x} \\ \dot{y} \end{bmatrix}_p = \begin{bmatrix} (E_R - E_L)/(2W) \\ (E_R + E_L)/2 \\ E_a + D\dot{\theta}_p \end{bmatrix} \quad (29)$$

W is the distance between a vertical encoder and the center vertical axis. The left and right encoders are symmetric to the center vertical axis. D is the distance between the horizontal encoder and the center horizontal axis. Refer to Figure 5. By rearranging (29), we can obtain:

$$\begin{bmatrix} E_L \\ E_R \\ E_a \end{bmatrix} = \begin{bmatrix} \dot{x} - W\dot{\theta} \\ \dot{x} + W\dot{\theta} \\ \dot{y} - D\dot{\theta} \end{bmatrix}_p \quad (30)$$

Combining (28) and (30) gives:

$$h\left(\begin{bmatrix} x \\ y \\ \theta \\ \dot{x} \\ \dot{y} \\ \dot{\theta} \end{bmatrix}_p\right) = \begin{bmatrix} x \\ y \\ \theta \\ \dot{x} - W\dot{\theta} \\ \dot{x} + W\dot{\theta} \\ \dot{y} - D\dot{\theta} \end{bmatrix}_p \quad (31)$$

Noise $r_t$ has covariance matrix R:

$$R = diag([d_{var} \quad d_{var} \quad d_{var} \quad E_{L_{var}} \quad E_{R_{var}} \quad E_{a_{var}}])$$

$$d_{var} = 0.001 d_{image}^2 + 0.001$$

$$E_{i_{var}} = 0.002 E_i^2 + 0.001 \quad (32)$$

Similar to the prediction covariance matrix, the measurement covariance matrix is not obtained from measurements on a physical robot. Odometry measurement noise increases as the encoder spins faster, and landmark detection is noisier if the robot is further away from the image. The variable dimage is the distance between the robot and the detected image.

*2) State-to-measurement Jacobian*

We can use (30) to compute the Jacobian H. Our h function is linear, resulting in a constant Jacobian H.

$$H = \begin{bmatrix} 1 & 0 & 0 & 0 & 0 & 0 \\ 0 & 1 & 0 & 0 & 0 & 0 \\ 0 & 0 & 1 & 0 & 0 & 0 \\ 0 & 0 & 0 & 1 & 0 & -W \\ 0 & 0 & 0 & 1 & 0 & W \\ 0 & 0 & 0 & 0 & 1 & -D \end{bmatrix} \quad (33)$$

## 5. PURE PURSUIT PATH FOLLOWING

The path following algorithm being used to test localization accuracy is Pure Pursuit [7]. Pure Pursuit is relatively easy to implement and requires low computational power. It works by pursuing the furthest point on the path that is within a specified distance from the robot.

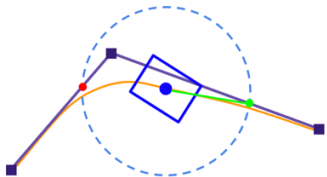

Figure 6: Pure Pursuit: The squares and the segments connecting them are the desired path. The blue rectangle is the robot and the blue point is its center. To follow the path, a circle with fixed radius is constructed around the robot, and the point to pursue is the intersection (green dot) of the circle with the path that is furthest along the path. If there are no intersections, the target point is just the nearest point on the path to the robot. The orange line is a hypothetical path the robot might take.

In order to pursue a target point, Pure Pursuit utilizes a Proportional Integral Derivative (PID) controller. A singular PID can only bring a scalar quantity to a target value, so in order to move a robot to a target position (three scalars), three PIDs are used in this case - one each for the x, y, and $\theta$ values. The control action $u_t$ a PID controller outputs is determined by this expression:

$$u_t = P \cdot p_t + I \cdot i_t + D \cdot d_t \quad (34)$$

Coefficients P, I, and D are constants that require tuning. Each update cycle, $p_t$ is set to the error, which is the target state minus the current state. $i_t$ is the sum of all the errors up to time t. $d_t$ is the change in error since the last update and is sometimes normalized over time. Let the current state be $s_t$ and the target state be $s_{target}$.

$$p_t = s_{target} - s_t \quad (35)$$

$$i_t = i_{t-1} + p_t \quad (36)$$

$$d_t = p_t - p_{t-1} \quad (37)$$

Eventually, the robot reaches the end of the path.

## 6. LOCALIZATION EXPERIMENTS

### 6.1. Accuracy Metric

Localization accuracy will be measured by the root mean squared error (RMSE) of specific state values for all time steps.

$(x, y)$ RMSE:    position translational error  
$x$ RMSE:    x position error  
$y$ RMSE:    y position error  
$\theta$ RMSE:    $\theta$ position error  
$\dot{x}$ RMSE    x velocity error  
$\dot{y}$ RMSE    y velocity error  
$\dot{\theta}$ RMSE    $\theta$ velocity error

For all EKF experiments, we will be using the same path (Figure 7). This path works well for testing localization because the path includes many unique types of movement to follow, such as straight lines, left/right turns, and elliptical curves.

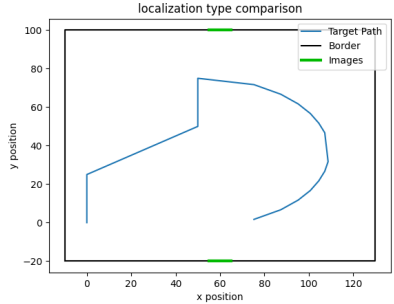

Figure 7: Path for Localization Test: The path starts at (0, 0) and ends at (75, 2). Images for landmark detection localization are located at the top and bottom of the border.

## 6.2. Velocity Accuracy Comparison

On a physical robot, system variance is often higher than measurement variance, so we imitate this in our simulation (Figure 8).

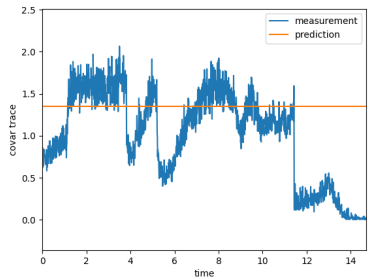

Figure 8: Covariance Comparison: The sum of the velocity variances of prediction covariance matrix Q and the measurement covariance matrix R are compared for the path in Figure 7. Since the prediction should be less accurate than the measurement, it has a higher variance most of the time.

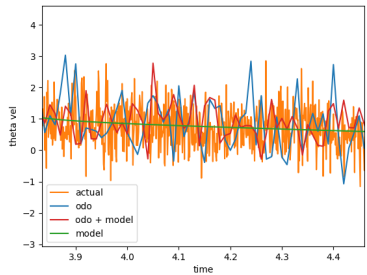

Figure 9: Velocity Fusion: The estimation from combining the model and the odometry measurement is more accurate than just using one of them alone.

To show velocity estimation (Figure 9) is better both prediction and measurement, we run 100 simulations using each of three localization types and find the average velocity RMSE. The localization types are using only prediction (model), using only measurement (odo), and using both prediction and measurement with an EKF.

The table above shows that fusing together the model with odometry with an EKF has the lowest RMSE for all three velocity states. The position RMSE with an EKF might be high because neither the prediction nor the measurement directly estimates the position. This makes it so slight angle error can lead to significant translational error.

| Localization without Camera | | | |
|---|---|---|---|
| type | model | odo | model + odo |
| $(x, y)$ RMSE | 2.2432 | 2.9770 | 2.8704 |
| $\theta$ RMSE | 0.1179 | 0.1903 | 0.1864 |
| $\dot{x}$ RMSE | 0.6692 | 0.4578 | 0.3589 |
| $\dot{y}$ RMSE | 0.6717 | 0.4120 | 0.3271 |
| $\dot{\theta}$ RMSE | 0.6698 | 0.2286 | 0.1996 |

## 6.3. Position Accuracy Comparison

Position correction shown in Figure 10, which is a result of data fusion, is both a benefit and drawback to the EKF. It is a benefit because the robot can more accurately localize. It is a drawback because the estimated position will suddenly shift, which can lead to jerky path-following movements.

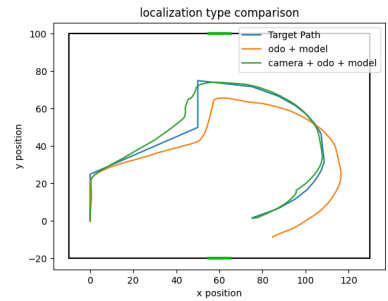

Figure 10: Data Fusion: When a landmark isn't in sight, the robot is able to use odometry to localize, but over time, the position error accumulates. Once a landmark is back in view, the position corrects itself.

The following RMSE values are obtained from the same 100 trials as earlier:

| Localization with Camera | | |
|---|---|---|
| type | odo + model | odo + model + camera |
| $(x, y)$ RMSE | 2.8704 | 1.2140 |
| $x$ RMSE | 6.4336 | 1.6400 |
| $y$ RMSE | 8.0909 | 1.8259 |
| $\theta$ RMSE | 0.1864 | 0.0956 |

## 6.4. Localization for FTC

In the FTC competition, the most common type of autonomous path is consistently and accurately cycling between two points. In the 2021-2022 FTC Freight Frenzy game (Figure 1 and 11) the robot needs to travel between the warehouse and ship hub while avoiding the barriers.

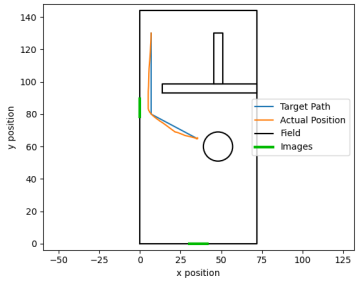

Figure 11: Cycle Path: The circle is the shipping hub and the top left corner is the warehouse. Landmark images are located on the left and bottom of the field (green). The robot is following the path using camera + odo + model for localization.

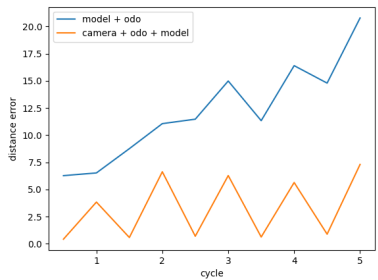

Figure 12: A "cycle" refers to the robot making its way from the warehouse to the shipping hub and back to the warehouse. The graphed data is an average of 10 trials.

Without using landmark detection localization, the accuracy of odometry will drift over time, leading to error accumulation in the long run as shown in Figure 12. The accuracy of odometry may not be realistic due to the 20-inch error accumulation in just 5 cycles, but error accumulation is still a serious issue to physical robots.

### 6.5. EKF Drawbacks

When $\Delta t$ is large, the Jacobians start to break down. This is because Jacobians are linear approximations, meaning their accuracy deteriorates as $\Delta t$ increases. Furthermore, odometry accuracy worsens when $\Delta t$ increases. Refer to Figure 13.

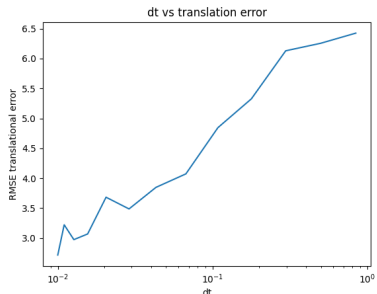

Figure 13: 30 simulations were run at each value of $\Delta t$, and the average (x, y) RMSE value was recorded. The time axis is logarithmic.

Another drawback to the EKF is if the system is highly nonlinear. The EKF works better when velocity is in body coordinates instead of global coordinates since body coordinates tend to be less nonlinear. For example, when moving in a circle at constant speed, global velocity is sinusoidal while body velocity is constant. Since the EKF works by performing many linear approximations, having a highly nonlinear velocity state leads to less accurate localization.

### 7. CONCLUSION

We successfully applied an Extended Kalman Filter to "fuse" odometry and computer vision landmark measurements. The EKF definitely provides good results for a mecanum drivetrain, as quantified by the decrease in RMSE when data is fused, and can be applied to the FTC competition. It greatly increases localization accuracy while not requiring much computation power.

The next steps are to implement an EKF on a physical robot with a more realistic model that includes acceleration and other factors such as friction and motor voltages. A more accurate physics model can be obtained by training a machine learning model with human driving. Another state estimation algorithm to look into is the particle filter, as it can perform state estimation on systems with non-Gaussian noise.